\documentclass[12pt,preprint]{aastex}

\shorttitle{Mason E.}
\shortauthors{V893 Sco}

\begin{document}

\title{High spectral resolution time-resolved optical 
spectroscopy of V893~Sco\footnote{Based on observations obtained at ESO La~Silla 
(ESO Proposal 64.H-0028(A))}}

\author{E. Mason}
\affil{Physics \& Astronomy Department, University of Wyoming,
    Laramie, WY 82071}
\and
\affil{Astrophysics Group, Planetary Science Institute, Tucson, AZ 85705}
\email{elena@psi.edu}

\author{W. Skidmore}
\affil{Physics \& Astronomy, University of St. Andrews, Scotland, UK}

\author{S.B. Howell}
\affil{Astrophysics Group, Planetary Science Institute, Tucson, AZ 85705}

\author{R.E. Mennickent}
\affil{Departamento de Fisica, Universidad de Concepcion, Chile}

\begin{abstract}
We present high resolution time-resolved optical spectra of the high 
inclination short orbital period dwarf nova V893~Sco. We performed spectral 
analysis  through radial velocity measurements, Doppler mapping, and ratioed 
Doppler maps. Our results indicate that V893~Sco's accretion disk is dissimilar to WZ~Sge's accretion disk, and does not fit any of the current accretion disk models. We derive the system parameters M$_1$ and $i$, and present evidence for V893~Sco as a very young cataclysmic variable and an ER~UMa star. We 
advance the hypothesis that all ER~UMa stars may be newly formed cataclysmic 
variables.
\end{abstract}

\keywords{Cataclysmic variables: dwarf novae, V893 Sco}

\section{Introduction}

Emission lines in high inclination short orbital period dwarf novae consist of strong hot spot emission superimposed on a double peaked profile from the 
accretion disk \citep{warner}. The hot spot emission, because of the binary 
system orbital motion, produces the characteristic S-wave motion seen in  
trailed spectrograms \citep{honey87}. 
Doppler maps of dwarf novae emission lines show clear indications of both the 
hot spot and the accretion disk emission, the latter producing a ring-shaped 
bright region in velocity space \citep{kait94}. Emission lines from the 
accretion disk imply that part of the accretion disk gas is optically thin and has low density \citep{kait94,will80,tyl81}. Optically thin, low gas density 
accretion disks are predicted in short orbital period dwarf novae, due to their low mass transfer rate \citep[$\dot{M} \lesssim 6.3\times 10^{14}$ gr  
sec$^{-1}$]{HSC95,hrp97}. 

Photometric and spectroscopic studies of short orbital period dwarf novae at 
quiescence show accretion disks which do not fit in the standard optically 
thick $\alpha$-disk model (e.g. WZ~Sge Mason et al. 2000, Skidmore et al. 2000; Z~Cha and OY~Car Wood 1990). In particular, recent works on WZ~Sge 
\citep{ele00,skid00} have shown evidence for an optically thin accretion disk 
in the emission lines, and an optically thick hot spot elongated along the 
stream trajectory. Observational testing is fundamental to prove or disprove 
current accretion disk models and evolutionary theories. Thus, the authors have undertaken a project aimed at the study of the accretion disks in quiescent, short orbital period dwarf novae, through time resolved spectroscopy. V893~Sco was selected because it is a high orbital inclination system showing hot spot eclipse \citep{bruch00} and strong double peaked Balmer emission lines similar to WZ~Sge. 

We describe our new observations of V893~Sco and the data reduction in Section 2. Results from trailed spectra, Doppler  and ratioed Doppler maps, and radial velocity curves are presented in Section \ref{sec3}. Section \ref{sec4} 
discusses our understanding of V893~Sco's accretion disk structure and compares it with that of WZ~Sge, a similar looking short orbital period dwarf nova. The present evolutionary status of V893~Sco is also discussed. Section \ref{sec5} summarizes our results.  

\section{Observation and Data Reduction} \label{sec2}

Observations  were carried out on 2000 March 3 and 4 UT on the NTT 
telescope in La~Silla. EMMI\footnote{EMMI is a multi-mode instrument for optical spectroscopy in low, medium, and high spectral resolution, and optical wide field imaging. EMMI has two arms, one optimized for the Blue and one for the Red wavelength range of the optical wave-band.} was used in medium resolution mode, with gratings Grat$\#$3 and Grat$\#$6 mounted in the Blue and the Red arm, respectively. Observations were performed simultaneously by the use of a 
dichroic with central wavelength 450 nm. A detailed log of the observations is summarized in Table \ref{tbl1}. 

Both nights were photometric and the seeing was always $\lesssim 0.8-0.9''$, 
thus, the slit aperture was set to $0.9''$. Each night the standard star Feige 56 was observed before V893~Sco for flux calibration. HeAr and ThAr arc 
exposures were taken only at the beginning and the end of the object 
observations and used for wavelength calibration of the Blue and the Red arm,  respectively. 

\begin{figure}
\epsscale{0.8}
\plotone{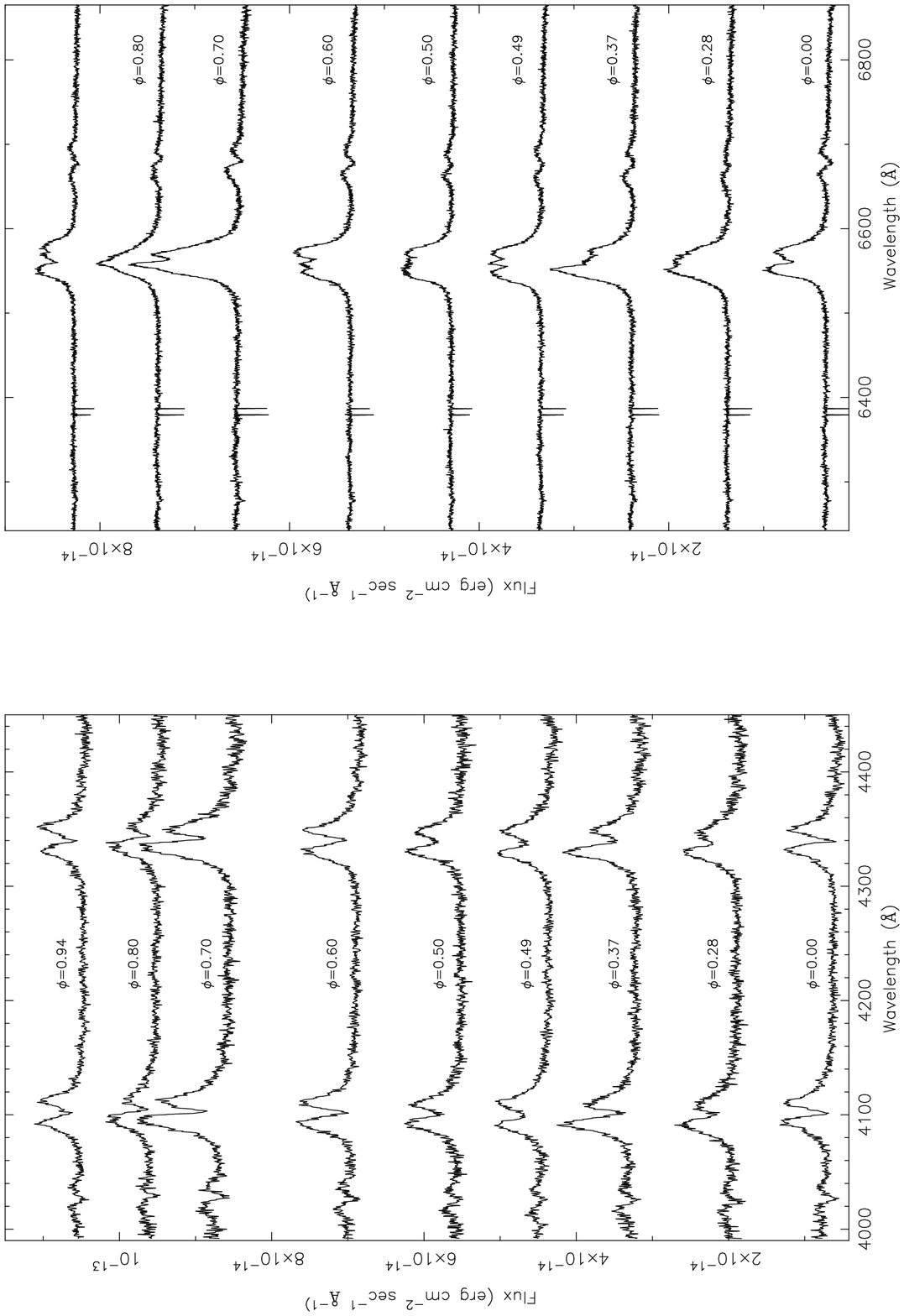}
\caption{Sample of Blue and Red Arm spectra showing, respectively, HeI 4026.3 H$\delta$ and H$\gamma$, and H$\alpha$ and HeI $\lambda$6678.1. Each spectrum is offset from the one below by $1.0\times10.^{-14}$ erg cm$^{-2}$ sec$^{-1} \AA^{-1}$ and $1.25\times10.^{-14}$ erg cm$^{-2}$ sec$^{-1} \AA^{-1}$ in the red and blue arm respectively. Narrow ``absorption lines'' in the red arm spectra are bad pixels. 
\label{fig1}}
\end{figure}

Data reduction was performed by one of us (E.M.) using the standard 
IRAF routines {\it ccdred} and {\it onedspec}. Each frame was bias subtracted 
and flat field corrected using exposures taken on the morning following each 
observing night. Wavelength calibration was performed by extracting the matching arc for each spectrum and by fitting the arc emission lines with a spline-3 of 1$^{st}$ or 2$^{nd}$ order in both arms. The number of fitted lines varied between 55-60 in the blue and 15-20 in the red; both fits were carried out 
keeping the residuals of the fitting function $\leqslant$ 10-12 Km sec$^{-1}$, i.e. $<0.25$ \AA/pixel. A sample of the reduced and phase folded spectra is 
shown in Figure \ref{fig1}

\section{The emission lines analysis} \label{sec3}

\subsection{Trailed spectra and Doppler maps}\label{sec31}

In V893~Sco, the emission line profile evolution through an orbit (see Figure 
\ref{fig1}), appears quite confusing at first glance. The three Balmer lines 
show different profiles and profile evolution. For example, the H$\alpha$ line shows multi-peaked profiles which are never observed in the blue arm spectra and  not explained by the standard S-wave motion. The two HeI emission lines 
$\lambda\lambda$4026.3 and 6678.1 seem to follow the profile evolution shown by the H$\gamma$ and H$\delta$ lines.

\begin{figure}
\epsscale{0.9}
\plotone{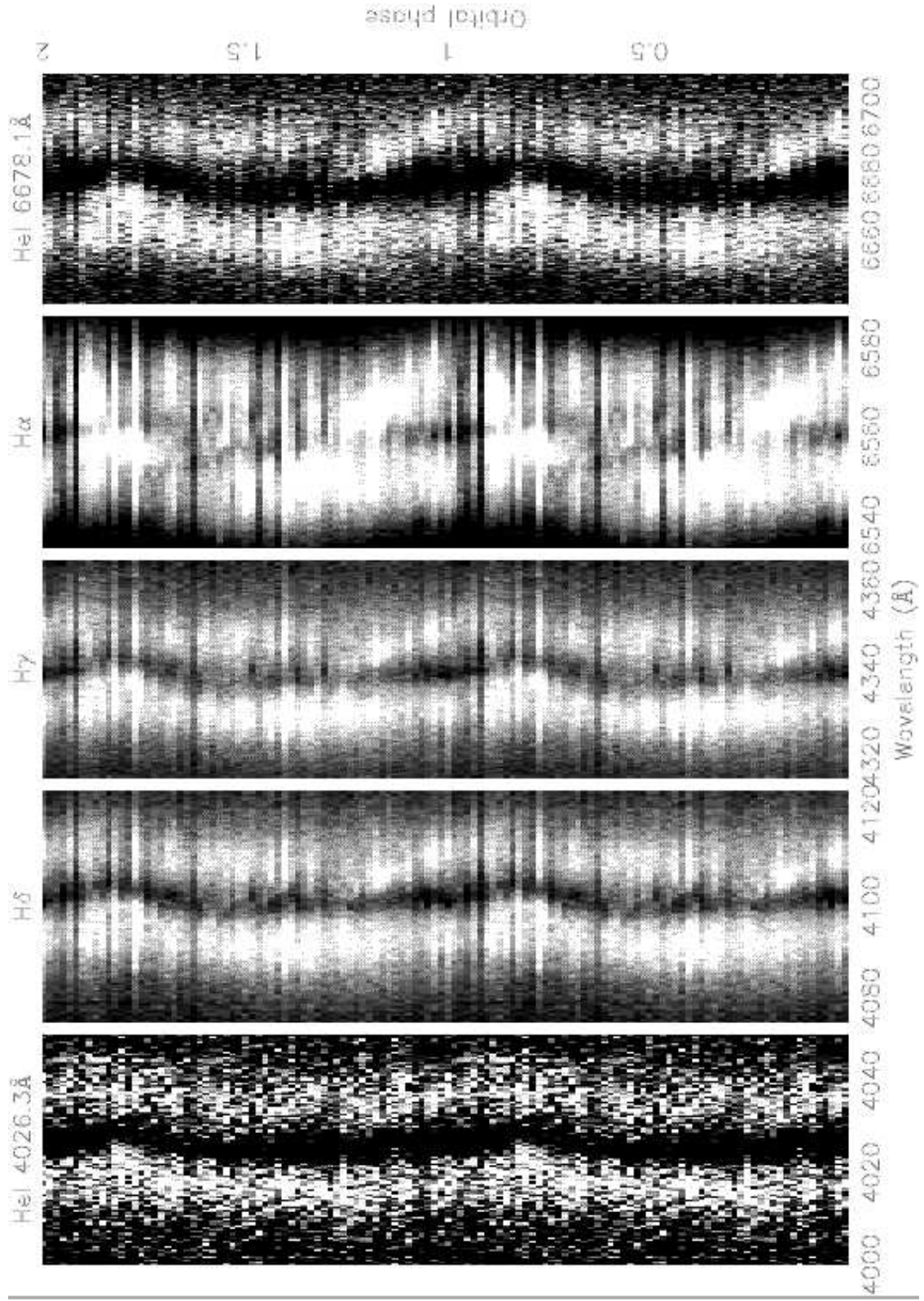}
\caption{Trailed spectrograms of the Balmer lines (H$\alpha$, 
H$\gamma$, and H$\delta$), and the HeI lines $\lambda\lambda$4026.3, 6678.1. 
 \label{fig2}}
\end{figure}

Trailed spectrograms (Figure \ref{fig2}) for the five emission lines show a lack of the typical S-wave component, and and the V/R ratio is $\gtrsim1$ at all phases. 
Since the lack of the S-wave component cannot be due to the time resolution of our spectra, the following two physical interpretations can be advanced: {\it 1)} there is not significant hot spot emission, {\it 2)} the accretion disk 
gas is optically thick in the emission lines, thus, the line emissivity is 
strongly anisotropic. The line photons emerge more readily in directions along which the Keplerian velocity field provides a large Doppler shifted gradient 
\citep{hm86}. The above two hypotheses are not mutually exclusive. 

Doppler maps of the Balmer lines have been generated by using  the back 
projection algorithm in the data reduction package {\it molly} (written by 
T.~Marsh, see Horne 1991 for details of the back projection technique). The 
spectra were continuum subtracted (the continuum was fit by a first order 
spline3) and Fourier Filtered before back projection (both blue and red arm 
spectra). Phasing of the spectra was made according to the photometric ephemeris given in \citet{bruch00}. Spectra around phase $\sim0$ were not excluded as we verified that the emission line fluxes do not show any eclipse effects. 

\begin{figure}
\epsscale{1.}
\plotone{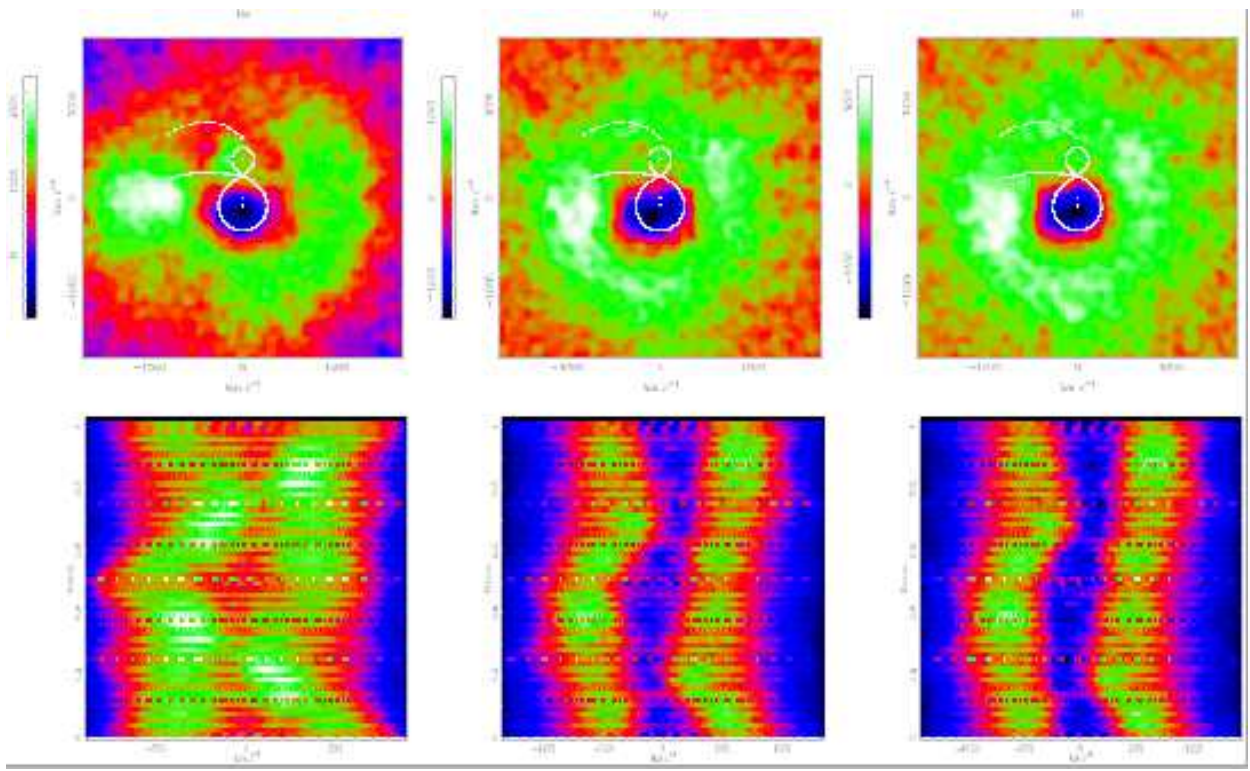}
\caption{Top: Doppler maps of the three Balmer lines. Filtering of the spectra was performed with a FWHM of 0.5 in all three cases; velocity-binning was made at the rest wavelength for each line with the velocity dispersion corresponding to a 0.39\AA/pixel resolution. Input $\gamma$ velocity was different for each Doppler map, and corresponds to the values determined for each radial velocity curve (see Table \ref{tbl3}). Bottom: Trailed spectrograms reconstructed by inverting the back projection process. By comparison with the trailed spectra in Figure~2 it is evident that some smoothing (in the time/phase direction) has occurred as expected. However, the main features characterizing the input data have been reproduced in the reconstruction.  
\label{fig3}}
\end{figure}

Figure \ref{fig3} shows Doppler maps and reconstructed trailed spectrograms for the three Balmer lines. All three Doppler maps show a relatively strong emission at roughly the expected hot spot position. Indeed, trailed spectra in Figure~2 show enhanced emission in the Blue and Red accretion disk peak in the phase ranges 0.1-0.25 and 0.4-0.55\footnote{In the case of the H$\alpha$ emission line also the phase range 0.7-0.85 contributes to the hot spot emission in the Doppler map.}, respectively, which may be evidence of some hot spot emission. The reconstructed trailed spectra show that such enhanced emission produces the hot spot region in the back projected Doppler maps. However, the hot spot emission is stronger than the accretion disk emission only by, at most, 20\% and $\lesssim$10\% in the H$\alpha$ and H$\gamma$ and H$\delta$ lines, respectively. 

In the H$\alpha$ Doppler map, the hot spot emission is centered at $v_X\sim-800$ and $v_Y\sim0$ Km sec$^{-1}$ and is quite elongated toward negative $v_X$ values. Previous H$\alpha$ Doppler maps of V893~Sco \citep{mmk00} show the hot spot centered at about the same position, but different in both size and orientation. The accretion disk emission appears to arise mainly from the leading side of the disk, being strong between phases 0.15-0.5 and weak otherwise. Non-uniform accretion disk emission was also pointed out by \citet{mmk00}, however their Doppler maps show the accretion disk inhomogeneities to dramatically vary both in extension and position from night to night. Our H$\alpha$ Doppler map of V893~Sco provides evidence for emission from the secondary star being irradiated by the hot spot. Irradiation of the secondary star was noticed also by \citet{mmk00}, but they observed it to disappear during one of their six nights of observation, just in the pre-outburst stage. 
There is no evidence of irradiation in the H$\gamma$ and H$\delta$ Doppler maps. 
Doppler maps for H$\gamma$ and H$\delta$ show a quite extended hot spot region, centered slightly ahead of the H$\alpha$ hot spot emission and possibly 
elongated along the $v_Y$ axis. The hot spot peak intensity appears at 
$v_X\sim-800$, $v_Y\sim-200$, and $v_X\sim-800$, $v_Y\sim-400$ Km sec$^{-1}$ for H$\gamma$ and H$\delta$ respectively, with the H$\gamma$ hot spot being double peaked. 

\subsection{Radial Profiles, Ratioed Doppler Maps, Balmer Decrement, and 
Continuum Fitting} \label{sec32}

\begin{figure}
\epsscale{1}
\plotone{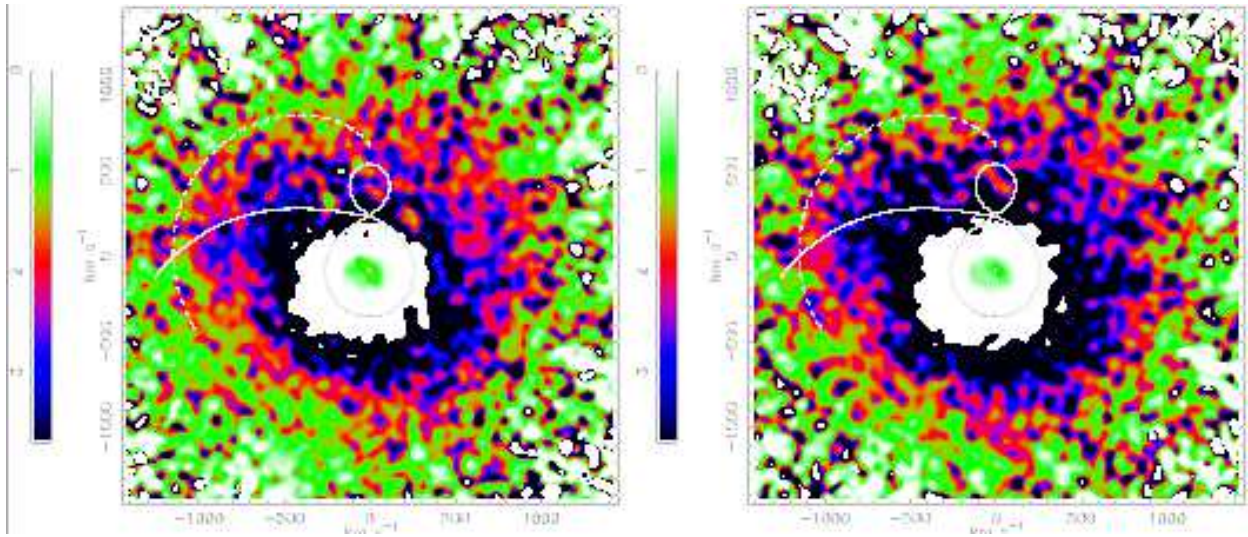}
\caption{Ratioed Doppler maps. Scale bars show the ratio of fluxes in 
frequency units. \label{fig4}}
\end{figure}

\begin{figure}
\epsscale{0.75}
\plotone{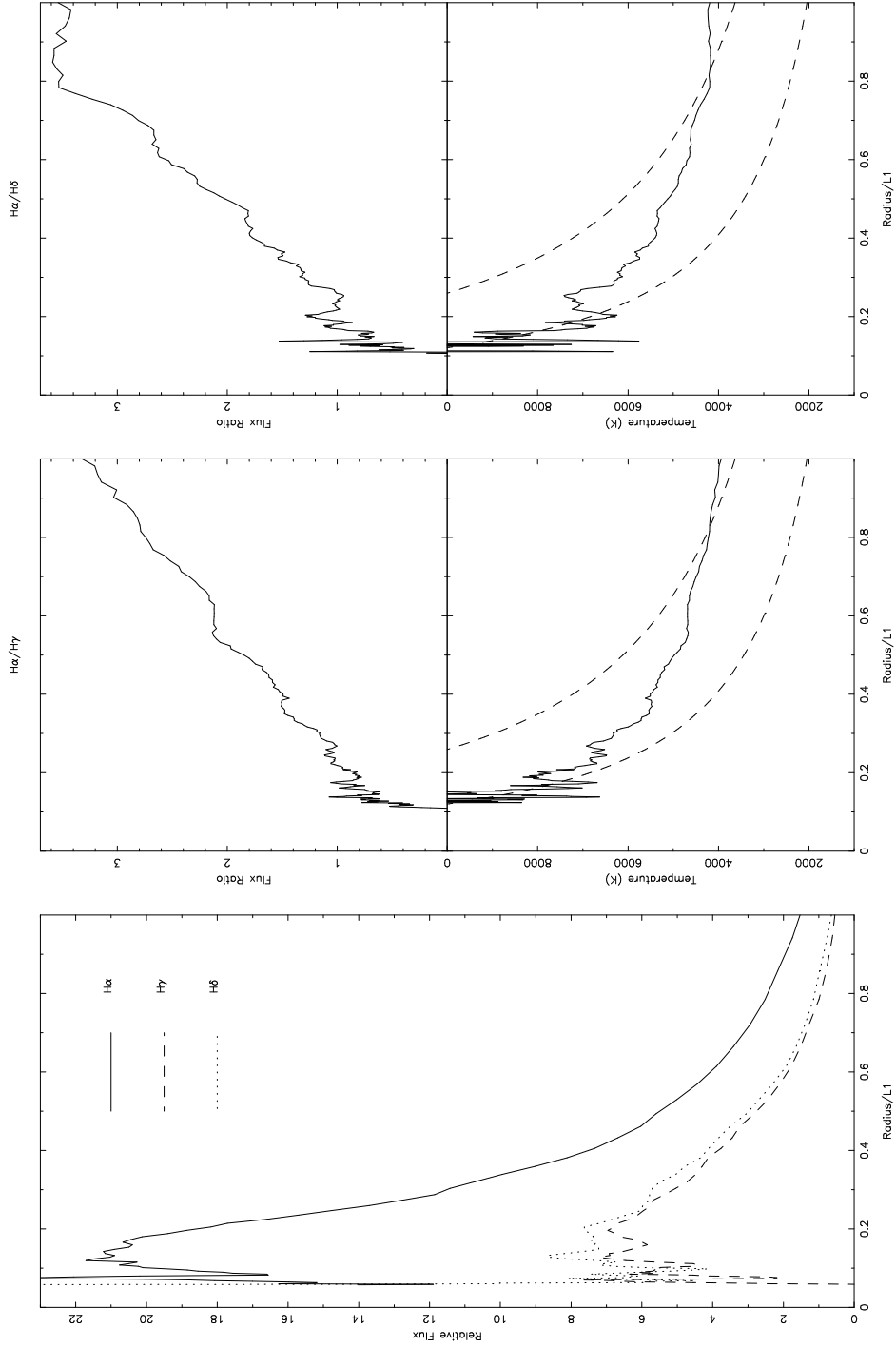}
\caption{Left panel: radial profile of the accretion disk flux in the 
three Balmer lines. Middle and Right panels: accretion disk flux ratio (top) and 
temperature (bottom) profiles derived from the ratioed Doppler maps for 
H$\alpha$/H$\gamma$ and H$\alpha$/H$\delta$, respectively. The temperature 
profile was  computed at each point with the temperature of a blackbody 
with the same Balmer decrement. These temperature profiles must be considered 
only as a diagnostic to compare with the predicted temperature profile,  
T$\varpropto r^{-3/4}$, for an optically thick $\alpha$-disk model (dashed 
line). The top and bottom dashed lines in each panel represent the temperature profile for an $\alpha$-disk with $\dot{M}$ of $\sim 10^{15}$ gr sec$^{-1}$ and $10^{-1}$ gr sec$^{-1}$, respectively. \label{fig5}}
\end{figure}

Ratioed Doppler maps\footnote{Details of producing Ratioed Doppler maps can be found in \citet{skid00}} (Figure \ref{fig4}) show quite uniform gas densities in the accretion disk and roughly equal flux ratio at the hot spot position and the accretion disk. 
Radial profiles of the accretion disk emitting regions (Figure \ref{fig5}, first panel) show the line flux to smoothly increase toward the center of the 
accretion disk in all three of the Balmer lines, while the radial profiles of ratioed Doppler maps increase with the radius (Figure 5, top panels 2 and 3). The derived temperature profiles (Figure 5, bottom panels 2 and 3) are flatter than in  an optically thick accretion disk. The innermost accretion disk region (0.1-0.2 R/L$_1$) which apparently matches the temperature profile of an optically thick $\alpha$-disk having $\dot{M}\sim 10^{16}$ gr sec$^{-1}$, is too noisy to allow any definitive conclusion.  

The Balmer decrements H$\alpha$/H$\gamma$ and H$\alpha$/H$\delta$ have been 
computed in three different ways: {\it 1)} by ratioing the emission line fluxes, {\it 2)} by ratioing the emission line peak intensity as described in 
\citet{ele00}, and {\it 3)} by averaging either azimuthally and radially 
different segments in the ratioed Doppler maps. In methods {\it 1)} and {\it 
2)}, orbital average values were used as we did not find phase dependent 
modulations. The Balmer decrement at the hot spot position determined in method {\it 3)} is dubious, as the hot spot emission overlaps the accretion disk one.  
However, despite the different assumptions involved, methods {\it 1)}, {\it 2)}, and {\it 3)} provide identical Balmer decrement values, within the uncertainties (see Table \ref{tbl2}). 

\begin{figure}
\epsscale{0.9}
\plotone{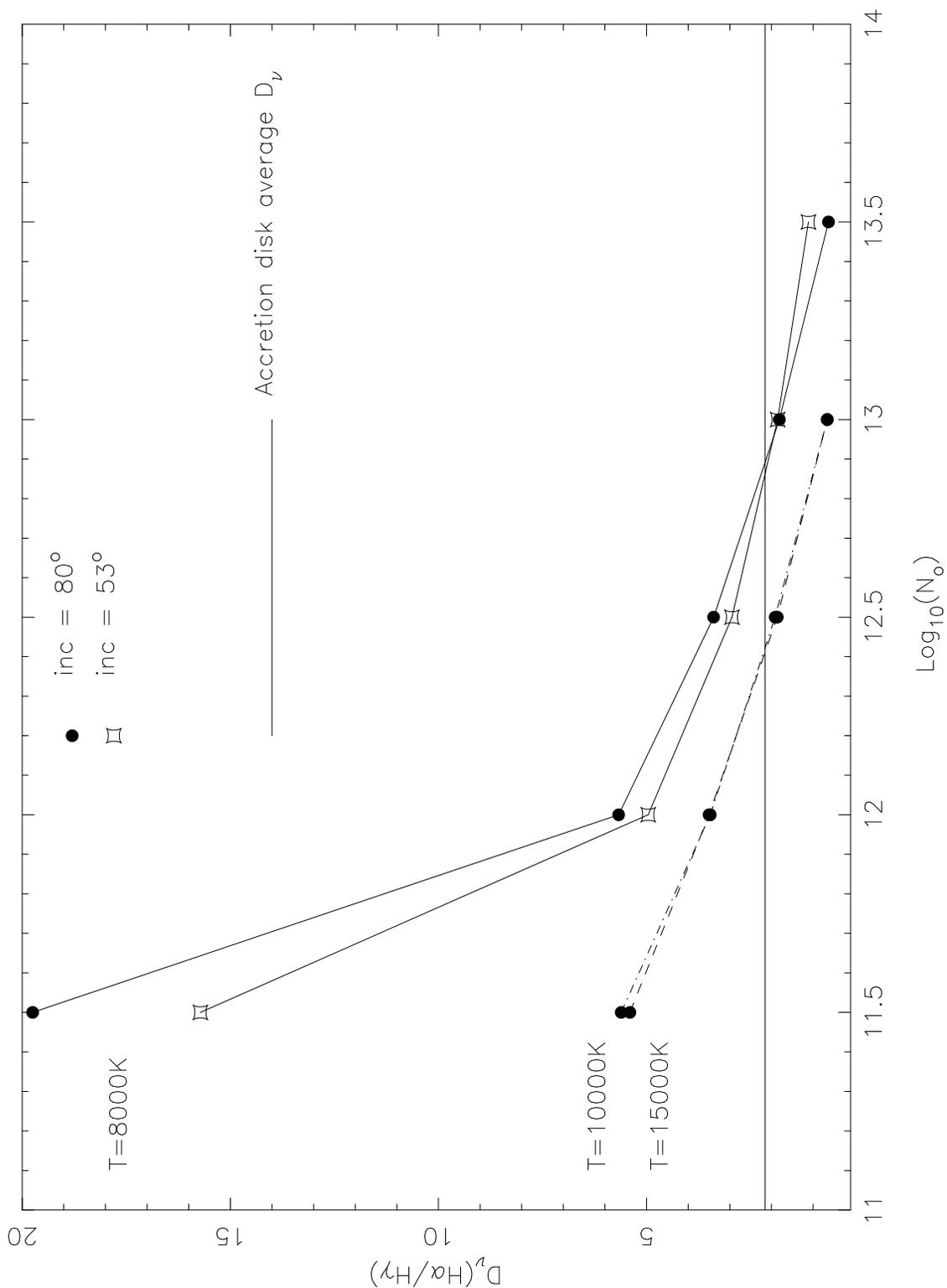}
\caption{The diagnostic diagram for H$\alpha$/H$\gamma$ for optically 
thin disks from models in \citet{will91}. Graph markers represent the Balmer 
decrement as function of the gas temperature and density. The horizontal line is our computed Balmer decrement for the gas within the accretion disk in V893~Sco. \label{fig6}}
\end{figure}

In order to derive information on the temperature (and possibly density) of the accretion disk line emitting region, we averaged the three values for 
each Balmer decrement $D_\nu(\frac{H\alpha}{H\gamma})$ and 
$D_\nu(\frac{H\alpha}{H\delta})$. We found 
$D_\nu(\frac{H\alpha}{H\gamma})=2.15\pm 0.29$ and 
$D_\nu(\frac{H\alpha}{H\delta})=2.47\pm 0.45$, which correspond either to a 
blackbody of temperature $\sim 4700\pm300$K\footnote{The blackbody 
temperature was computed including a thermal and/or turbulent broadening 
correcting factor of $\lambda_o^{-1}$ \citep{will80}. Neglecting the thermal 
broadening we find an upper limit of $\sim5700\pm300$ K for the accretion disk temperature in the optically thick case.}, or to an optically thin gas either of temperature T$\sim$8000K and density $\log_{10} N_o \sim 12.9$, or temperature T$\sim$10000-15000K and density $\log_{10} N_o \sim 12.4$ \citep[and Figure \ref{fig6}]{will91}. In the case of optically thick accretion disk 
emission lines, the hot spot gas cannot have the same Balmer decrement as the 
accretion disk, otherwise it implies a hot spot gas temperature of T$<5000$K, 
which is far too low for the gas impact region.  

We also investigated the continuum emission from both hot spot and accretion 
disk gas. We determined the hot spot continuum emission by subtracting the 
average of 5 spectra centered at phase $\sim$0 (hot spot eclipse) from the 
average of 5 spectra around phase $\sim$0.8 (hot spot facing the observer), and compared the result with a blackbody. The red and blue spectral differences 
could not be fit with a single blackbody function. We found a 8550$\pm$150 K and a 10450$\pm$50 K blackbody to match the continuum slope in the red and blue arm, respectively, possibly indicating that there is a temperature structure in the optically thick hot spot continuum emission. 
The accretion disk continuum was derived by fitting the continuum in the blue 
and red arm spectra averaged over both nights. We checked to see if the 
continuum slope matches an optically thick $\alpha$-disk emission, but found it to be shallower than $F_\lambda\varpropto\lambda^{-7/3}$, matching a best fit blackbody of 11750$\pm$250 K and 6400$\pm$50 K in the blue and red arm, 
respectively. The hotter blue continuum possibly indicates a higher gas 
temperature in the inner accretion disk. 

\subsection{Radial Velocity Curves} \label{sec33}

Radial velocity measurements were made on each original spectrum following the 
Pogson-like method described in \citet{ele00}. 
We measured the accretion disk emission line wings by positioning the cursor at different levels above the continuum up to $\sim30\%$ the average line 
intensity $I$. We verified the wing measurements at intensity levels $\lesssim 0.3\times I$ to not be biased by the hot spot S-wave motion, which is expected to blue-shift the blue wing and red-shift the red wing at phase 0.55 and 0.05 respectively. We observe such a hot spot bias to affect wings measurements at intensity levels $\gtrsim 0.5\times I$. However, we still derived different 
radial velocity curve fitting parameters depending on the measured emission 
line. Thus, bias other than the hot spot, must affect our radial velocity 
measurements. 

\begin{figure}
\epsscale{0.8}
\plotone{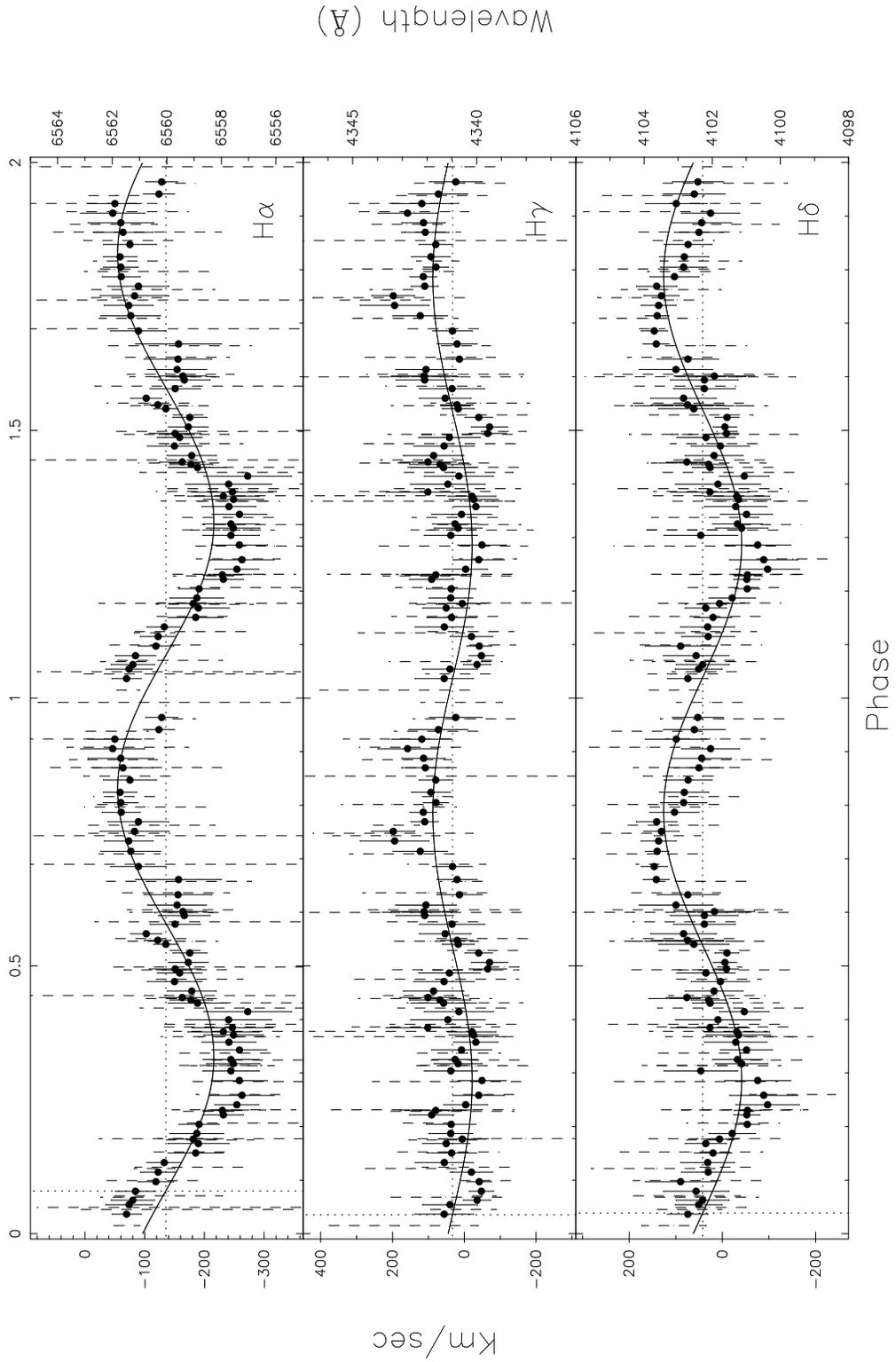}
\caption{Radial velocity measurements along with their best fitting 
sinusoidal curves. Filled circles and solid error-bars represent boxcar averaged points. Dashed lines show the unaveraged data. \label{fig7}}
\end{figure}

In order to reduce the uncertainties of the radial velocities, we applied a 
running-boxcar to our measurements before sine fitting the radial velocity 
curves. Boxcar smoothing of the radial velocity measurements was preferred to 
spectra-binning before radial velocity measurements as it does not reduce 
the phase resolution of the data set. The three smoothed radial velocity curves were fitted with the fitting function as in formula (1) of \citet{ele00}. The best fitting parameters are summarized in Table \ref{tbl3}, while the radial velocity curves and their correspondent fits are plotted in Figure \ref{fig7}. The HeI lines were too weak to allow accurate radial velocity measurements. 

The phase of the red-to-blue crossing, $\phi_o$ (Table \ref{tbl3}), corresponds to the phase offset between the spectroscopic and the photometric time for 
the secondary star inferior conjunction, and is believed to be proportional to the relative intensity of the hot spot emission with respect to the accretion disk emission, and to the hot spot bias in the radial velocity measurements.  In the V893~Sco radial velocity curves, the phase offset is small in all three of the Balmer lines, with an average value of $\overline{\phi_o}=0.0616\pm0.0090$. 
We also found the phase offset not to be proportional to the excitation 
potential energy of the H emission lines, in contrast with WZ~Sge and VY~Aqr 
\citep[and references therein]{ele00}. 

\section{Discussion} \label{sec4}

V893~Sco and WZ~Sge are high inclination, short orbital period dwarf novae, both showing a hot spot eclipse in photometric observations and strong double peaked Balmer emission lines in their optical spectra. However, despite these 
similarities, our spectral analysis applied to V893~Sco and WZ~Sge \citep{ele00} provided contrasting results. 

The emission lines in WZ~Sge are dominated by the hot spot emission which 
heavily affects the line profile evolution throughout the orbital period. In 
V893~Sco, the emission lines does not show strong evidence of hot spot emission in any of our spectra, with the exception of H$\alpha$. H$\gamma$ and H$\delta$, and the two HeI lines $\lambda\lambda$4026.3 and 6678.1, show V-shaped profiles and V/R ratios $\gtrsim$1 throughout the orbit. Trailed spectrograms of WZ~Sge are dominated by the S-wave component, while in V893~Sco, the trailed spectrograms show a complete lack of such a component.

Doppler maps of both WZ~Sge and V893~Sco show asymmetric accretion disk 
emission. However, in WZ~Sge, the disk emission appears weak and is superposed by an extremely strong hot spot emission, while the V893~Sco Doppler maps there is the same Balmer decrement at the hot spot position and in the accretion disk gas. Radial profiles of V893~Sco Doppler maps show increasing flux toward the center of the accretion disk, while in the case of WZ~Sge the emission line flux is observed to increase outward.  

Ratioed Doppler maps of WZ~Sge show a statistically significant larger Balmer 
decrement in the accretion disk than in the hot spot, while in V893~Sco ratioed Doppler maps the Balmer decrement at the hot spot position is larger than or equal to that in the accretion disk gas. Radial profiles of the ratioed Doppler maps and the derived temperature profiles both in the case of V893~Sco and of WZ~Sge do not match the profile predicted by an optically thick $\alpha$-disk model. Both systems have flux ratio increasing outward and a quite flat temperature profile. 

The radial velocity curves yield different values for the system parameters 
K$_1$, $\gamma$, and $\phi_o$, depending on the accretion disk emission line 
measured. This is true for both WZ~Sge and V893~Sco. However, in the case of 
V893~Sco, the derived phase offsets are small and not proportional to the 
excitation potential energy of the emission lines. 

In V893~Sco, the V-shaped profile of the emission lines, and the V/R ratios 
$\gtrsim$1, are consistent with saturated emission lines \citep{hm86}, and 
anisotropic turbulence with positive correlation in the vertical and azimuthal velocity components \citep{hr95}. Optically thick emission lines also explain the lack of S-wave components in trailed spectrograms. 
However, the accretion disk gas temperature of $\sim 4700$K in the line forming region does not match with any of the current accretion disk models at the mass transfer rate expected in V893~Sco \citep[ and see below]{osaki96}. 
\citet{tyl81} derives gas temperatures in the range 4000-5000K, only in the case of either very low mass transfer rates ($\dot{M}\sim10^{13}$ gr sec$^{-1}$), or high mass transfer rate and extended accretion disks ($\dot{M}\sim10^{17}$ gr sec$^{-1}$, $R_d>4\times10^{10}$ cm). \citet{will80} predicts a lower limit of 6000K for the accretion disk line forming region at any mass transfer rate. Both \citet{tyl81} and \citet{will80} predict emission lines to form in the outer accretion disk, while we observe the emission line flux to mostly arise from the inner accretion region.  

Our observed emission line flux radial profiles fit the optically thin line 
hypothesis where the line forming region is a hot corona or a gas layer above 
the accretion disk. Within such a hypothesis, we would expect the accretion disk continuum flux to be optically thick to explain the lack of S-wave component in trailed spectra and the photometric light curve \citep{bruch00}. However, our continuum analysis was not conclusive and two different blackbodies were found to match our spectra, rather than the expected flux distribution of the 
optically thick  $\alpha$-disk model. Furthermore, the optically thin emission line model does not explain the V-shaped line profiles. 
The observed V/R ratio $\geqslant$1 may be explained by a bipolar wind flowing perpendicularly to the disk plane with an outflow velocity of $v_{wind}\sim2000$ km sec$^{-1}$. However, wind in non magnetic cataclysmic variables is typically observed in the UV band and either in outbursting dwarf novae and in nova-likes (Warner 1995). The only systems where wind has been detected in the optical are BZ~Cam (Ringwald and Naylor 1998) and V592~Cas (Huber et al. 1998).

In order to better understand the physics of the ongoing accretion process, the binary system evolutionary status, and possibly constrain the accretion disk, we performed the following computations. We determined the secondary star mass, M$_2$, and radius, R$_2$, for a 1.82 hr orbital period system (see Howell and Skidmore, 2000). We find M$_2=0.175M_\odot$ and R$_2=0.196M_\odot$. We then used this R$_2$, with a first guess for M$_1$, R$_1$, $r_d$, and the hot spot radius \citep[$r_S=3.1\times 10^8$ cm,]{smak93} and azimuth ($\alpha=30^o$), to geometrically constrain the orbital inclination $i$. Applying formula 2.79 in \citet{warner} with our K$_1$(H$\alpha$) determination, we computed the white dwarf mass M$_1$. We iterated the computation of $i$ and M$_1$ until 
convergence, keeping constant all the other parameters. The results are 
$i=72.5^o$ and M$_1$=0.89$M_\odot$. 

We measured the emission line peak separation in the spectra averaged over both nights (see Table~4), to determine the accretion disk radius via the Equation (2) in 
\citet{ele00}. We found the accretion disk to extend to $\sim 0.5 R_{L_1}$ in 
H$\gamma$ and H$\delta$ emission lines, and to exceed the primary Roche lobe in H$\alpha$. We believe the H$\alpha$ average profile to be biased by both the hot spot and the secondary star emissions which reduce the accretion disk line peak separation. 

\begin{figure}
\epsscale{0.9}
\plotone{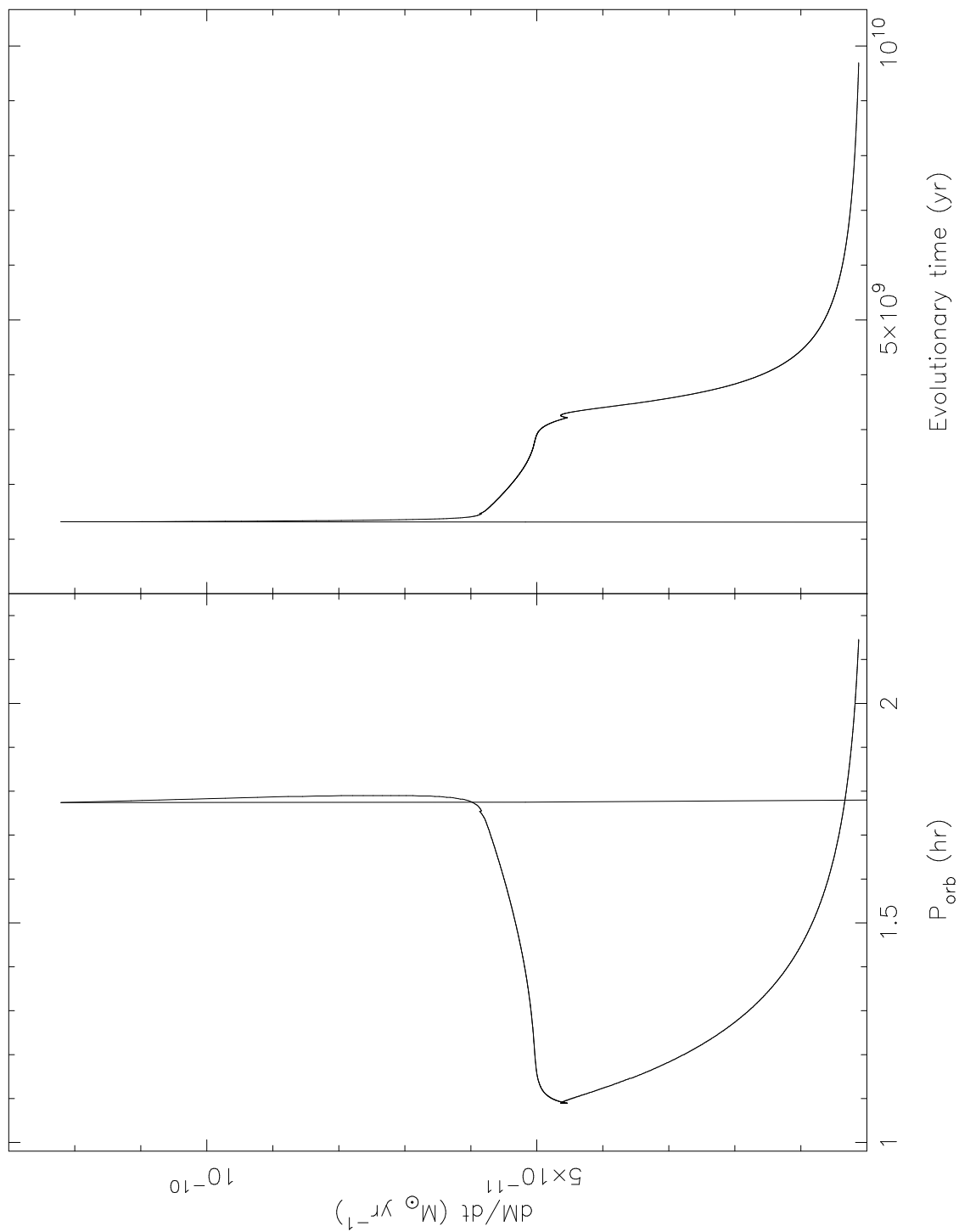}
\caption{Left panel: the evolutionary track of the mass transfer rate 
$\dot{M}$ vs the orbital period. Right panel: the evolutionary track for the 
mass transfer vs the evolutionary time (time zero is the birth of the system as two star out of the common envelope). Both plots show how the initial mass 
transfer rate for a cataclysmic variable of masses M$_1=0.89$M$_\odot$, 
M$_2$=0.175M$_\odot$, and $P_{orb}=1.82$ hr, is much higher than normal for a 
short time period. \label{fig8}}
\end{figure}

We can calculate an evolutionary model for V893~Sco using the code described in \citet[P$_{orb}$=1.82 hr, M$_1$=0.89$M_\odot$, and M$_2$=0.175$M_\odot$]{h01}. 
Figure \ref{fig8} plots the model results for the mass transfer rate  both as a function of the orbital period (left panel) and of the ``evolutionary time'' (right panel). Figure \ref{fig8} indicates that for the determined parameters in V893~Sco, the system is likely to be a newly formed cataclysmic variable having just begun mass transfer. This ``initial contact'' evolutionary phase is relatively short ($\lesssim 10^8$ yr), and characterized by an increased  
$\dot{M}$ due to the sudden onset of  mass transfer from the secondary star as it tries to adjust itself toward a new equilibrium \citep{h01}. 

Based on the orbital period and optical appearance, V893~Sco is believed to be an SU~UMa star, although  none of its observed outbursts is time-resolved enough to provide evidence for the occurrence of super-outbursts with super-humps. However, the VSNET database shows outbursts every $\sim$30 days spaced by  
possible ``mini''-outbursts ($\lesssim$1 mag) every few days. Such a photometric signatures characterizes V893~Sco as a high mass transfer rate system belonging to the ER~UMa stars class. The outburst behavior observed in V893~Sco, i.e. short recurrence time ($\lesssim 30$ days) and small outburst amplitude 
($\lesssim 2$ mags) is consistent with a cataclysmic variable having a high mass transfer rate and a hot accretion disk at or near constant outburst, similar to the nova-likes. 

Given the possible membership of V893~Sco as an ER~UMa star and its probable 
status as a young cataclysmic variable, we examined the evolutionary code also for the ER~UMa stars ER~UMa, V1159~Ori, and RZ~LMi. The masses of the two star components in each system were derived similarly to M$_1$ and M$_2$ in V893~Sco. 
However, given that none of the three ER~UMa stars present any eclipse or 
partial eclipse, the imposed  geometrical constraints were different. Formula 
2.79 in Warner (1995) was applied using as input parameters 
M$_1=\overline{M_1}=(1.4+M_2/0.25)/2$ (where M$_1=1.4M_\odot$ is the 
Chandraseckhar upper limit, and M$_1=M_2/0.25$ is the lower limit derived by the constraint $q<0.25$ in super-humps theory, Whitehurst 1988), $i=i_{max}/2$ 
(where $i_{max}$ is the average of the two upper limits for no eclipse derived for either a M$_1=1.4M_\odot$ and M$_1=M_2/0.25$), and $P_{orb}$ and K$_1$ 
determined by Thorstensen et al. (1997), and Szkody et al. (1996). In order to decrease the uncertainties on M$_1$, we then averaged the derived M$_1$ with results from observed white dwarf mass distributions in cataclysmic variables  \citep{sion99,web90}. The input parameters for our evolutionary models 
corresponding to each ER~UMa star are listed in Table \ref{tbl5}. The model 
results show all three of the ER~UMa stars to also be at the ``$\dot{M}$-spike'' during initial contact, thus being very young cataclysmic variables, similar to V893~Sco. We tested the validity of such a result by making the evolutionary models also for some SU~UMa stars with known M$_1$, M$_2$, and $P_{orb}$ 
\citep{rc98}.The model results were contrary to the observations, predicting 
incorrect orbital periods for the input masses and vice-versa. We believe that incorrect M$_1$ determinations may be the cause as they depend, by up to 70$\%$, on the input M$_2$ (from old, inaccurate M$_2$-$P_{orb}$ relations) in formula 2.79 in Warner (1995). The discrepant M$_1$ values derived for V893~Sco by \citet{mmk00} and us, arise from their use of an improper M$_2$-$P_{orb}$ 
relationship.    

\section{Summary $\&$ Conclusions} \label{sec5}

High resolution, time resolved spectroscopy of V893~Sco was analyzed and we 
found the following results: 
\begin{itemize}
\item Trailed spectra of the emission lines H$\alpha$, H$\gamma$, and 
H$\delta$, HeI $\lambda$4026.3, and HeI $\lambda$ 6678.1 were presented. They 
all show a lack of the S-wave component and V/R ratio $\gtrsim$1 (H$\alpha$ 
being the exception) across the whole orbit. Doppler maps of the three Balmer 
lines were presented. They show a non-uniform accretion disk with the hot spot emission only being $\sim20\%$ stronger than the accretion disk. Doppler maps and trailed spectra also show evidence for H$\alpha$ emission from the secondary star irradiated by the hot spot. 
\item The Balmer decrements, $D_\nu (\frac{H\alpha}{H\gamma})$ and $D_\nu 
(\frac{H\alpha}{H\delta})$, were determined by three different methods, all  
which provide similar values. Average Balmer decrements were used to determine the physical parameters of the accretion disk gas. Our computed values 
correspond to either an optically thin gas of T$\sim$8000-15000K or an optically thick gas of T$\sim$4700K. None of the current models describing emission lines from accretion disks completely satisfy the details of our results, thus, a definitive conclusion about the accretion disk gas temperature can not be 
reached in the present paper. 
\item Radial velocity curves for different emission lines yield inconsistent set of systemic parameters K$_1$, $\gamma$, and $\phi_o$ (as already found in 
WZ~Sge). However, the results are different with respect to WZ~Sge as we find 
only a small phase offset in V893~Sco and no linear-relationship between the 
orbital phase and the excitation potential energy for hydrogen. We conclude that the hot spot region is not elongated along the steam trajectory with gas 
temperature decreasing downstream. The hot spot line forming region in V893~Sco is consistent with the gas stream impacting on a denser accretion disk gas. 
\item The white dwarf mass was determined from the secondary star mass 
function and the measured K$_1$(H$\alpha$) amplitude. We find 
 M$_{WD}$=0.89M$_\odot$ and a mass ratio $q=0.19$. These system parameters 
match those of a newly formed cataclysmic variable in our evolutionary 
models. The high mass transfer rate predicted by either the model and the long term light curve of V893~Sco (VSNET database), together with the accretion disk properties we described in the previous section, imply that V893~Sco is likely to be an ER~UMa star. The evolutionary code of \citet{h01} also predicts three other ER~UMa stars to be newly formed cataclysmic variables. 
\end{itemize}

The two high inclination, short orbital period systems, V893~Sco and WZ~Sge, 
appear to be spectroscopically similar, yet are two completely different types of system. They not only differ in their accretion disk structure/physics 
as described in Section \ref{sec4} but also in their outburst behavior. WZ~Sge has the longest recurrence time ($\sim33$ years) and the largest outburst 
amplitude ($A\gtrsim 8$ mag); V893~Sco has the smallest outburst amplitude 
($A\sim$1.5-2.0 mag) and a short outburst recurrence time ($\lesssim$30 days). 
The derived properties for the accretion disks in these two stars appear 
consistent with current evolutionary theory. WZ~Sge has been shown to not fit 
the standard $\alpha$-disk model \citep{skid00,ele00}, is expected to have an
extremely low mass transfer rate \citep[$\dot{M}\leqslant 10^{15}$ gr 
sec$^{-1}$]{HSC95,osaki95,smak93}, and to be near or past the orbital period 
minimum \citep{HSC95,ciardi98}.

V893~Sco also does not fit the the standard $\alpha$-disk model, nor any of 
models accounting for at least a partly optically thin disk. We show it to be a high mass transfer rate, short orbital period dwarf novae with $\dot{M}$ in the range 10$^{15-16}$ gr sec$^{-1}$  and predict it to be an ER~UMa star, and a newly formed cataclysmic variable. The high mass transfer rate invoked to 
explain the outburst recurrence time would be a natural consequence of the 
sudden onset of the mass transfer process. However, time resolved light curves and long term monitoring are necessary to definitely prove or disprove our 
hypothesis that V893~Sco is an ER~UMa star, by detecting super-humps and 
confirming mini-outbursts. 

\acknowledgments
This research was partially supported by the NSF grant AST 98-10770 and from 
the University of Wyoming office of research. R.M. acknowledges support by
Fondecyt 1000324 and D.I. 99.11.28-1.

\clearpage

\clearpage

\begin{deluxetable}{ccccccc}
\tabletypesize{\scriptsize}
\tablecaption{Log of observations. \label{tbl1}}
\tablewidth{0pt}
\tablehead{
\colhead{Obs Date (UT)} & \colhead{$\lambda$-range (\AA)}   & \colhead{Em line}  
 &
\colhead{Disp (A/pix)} &
\colhead{Exp time (sec)}  & \colhead{$\#$ of Orbits} & \colhead{$\#$ of 
Spectra} 
}
\startdata
03/03/2000 & 6243-6865 & H$\alpha$ & 0.39 & 300 & 1.45 & 24 \\
03/03/2000 & 3990-4550 & H$\gamma$ + H$\delta$ & 0.39 & 300 & 1.45 & 24 \\
04/03/2000 & 6243-6865 & H$\alpha$ & 0.39 & 300 & 2.26 & 38  \\
04/03/2000 & 3990-4550 & H$\gamma$ + H$\delta$ & 0.39 & 300 & 2.26 & 38 \\
\enddata
\end{deluxetable}


\begin{deluxetable}{cccccc} 
\tabletypesize{\scriptsize}
\tablecolumns{7} 
\tablewidth{0pc} 
\tablecaption{Average Balmer decrements \label{tbl2}} 
\tablehead{ 
\colhead{} & \colhead{(method 1)} & \multicolumn{2}{c}{(method 2)} & 
  \multicolumn{2}{c}{(method 3)} \\
\colhead{$D_\nu$} &  \colhead{Emission line}  & 
\multicolumn{2}{c}{Peak Intensity} & \multicolumn{2}{c}{Ratio Maps Average 
Radial Profile} \\ 
\colhead{}   & \colhead{}    & \colhead{Disk} & 
\colhead{Hot Spot}    & \colhead{Disk}   & \colhead{Hot Spot}}
\startdata 
$\frac{H\alpha}{H\gamma}$ & 2.27$\pm$0.45  & 2.03$\pm$0.42 & 
	2.39$\pm$0.76 & 2.25$\pm$1.02 & 2.22$\pm$0.93 \\
$\frac{H\alpha}{H\delta}$ & 3.05$\pm$0.74 & 2.01$\pm$0.62 &
	2.34$\pm$0.75 & 2.68$\pm$1.34 & 2.61$\pm$1.29 \\
\enddata
\tablecomments{Balmer decrement values derived for the hot spot gas via 
method 3) (last column in the table) are not subtracted of the accretion disk 
contribution. See text for further explanations. }
\end{deluxetable}


\begin{deluxetable}{ccccc}
\tablecaption{Results from the radial velocity curve fits. \label{tbl3}}
\tablewidth{0pt}
\tablehead{
\colhead{Line} & \colhead{Central Wavelength}   & \colhead{V/R crossing}   &
\colhead{$K_1$ amplitude} &
\colhead{Systemic velocity ($\gamma$)}\\
\colhead{} & \colhead{\AA} & \colhead{$\phi_o$} & \colhead{km sec$^{-1}$} & 
\colhead{km sec$^{-1}$}   
}
\startdata
H$\alpha$ & 6560.03745 & 0.0798$\pm$0.012 & 80.6$\pm$7.5 & -135.3$\pm$4.8 \\
H$\gamma$ & 4340.97680 & 0.0354$\pm$0.027 & 54.6$\pm$8.6 & 32.9$\pm$7.7 \\
H$\delta$ & 4102.28000 & 0.0385$\pm$0.016 & 83.7$\pm$7.6 & 42.4$\pm$5.4 \\
\enddata
\end{deluxetable}


\begin{deluxetable}{cc}
\tabletypesize{\scriptsize}
\tablecaption{The Keplerian velocity as measured from averaged spectra. 
\label{tbl4}}
\tablewidth{0pt}
\tablehead{
\colhead{Emission Line} & \colhead{Half peak separation}\\
\colhead{ } & \colhead{(km sec$^{-1}$)} \\
}
\startdata
H$\alpha$ & 393.3$\pm$4.6 \\
H$\gamma$ & 606.2$\pm$6.9 \\
H$\delta$ & 649.2$\pm$7.3 \\
\enddata
\end{deluxetable}


\begin{deluxetable}{ccccccccccccc}
\tabletypesize{\scriptsize}
\tablecaption{The masses derived for the three ER~UMa stars. 
\label{tbl5}}
\tablewidth{0pt}
\tablehead{
\colhead{Star name}  & & & & & & \colhead{M$_2$ (M$_\odot$)} & & & & & & 
\colhead{M$_1$ (M$_\odot$)} \\
}
\startdata
ER~UMa  & & & & & & 0.14 & & & & & & 0.86$\pm$0.12\\
V1159~Ori & & & & & & 0.14 & & & & & & 0.86$\pm$0.12\\
RZ~LMi & & & & & & 0.12 &  & & & & & 0.85$\pm$0.10 \\
\enddata
\tablecomments{M$_1$ determined for RZ~LMi is an upper limit. 
\citet{setal96} could not fit any sine curve to their radial velocity curve.  
Thus, K$_1$=20 Km sec$^{-1}$ is estimated to 
be an upper limit for the white dwarf Keplerian velocity.}
\end{deluxetable}

\end{document}